\documentclass[10pt,superscriptaddress,nofootinbib,twocolumn,pra,aps]{revtex4-2}
\usepackage{amsthm,amsmath,amssymb}
\usepackage{mathrsfs}
\usepackage{amsthm}
\usepackage{amsmath}
\usepackage{amssymb}
\usepackage{leftindex}
\usepackage{graphicx}
\usepackage{epstopdf}
\usepackage{url}
\usepackage{bm}
\usepackage{ifthen}
\usepackage[usenames,dvipsnames]{color}
\usepackage{mathrsfs}
\usepackage[colorlinks=true,citecolor=blue,urlcolor=black]{hyperref}
\usepackage{subfigure}
\usepackage{multirow}
\usepackage{makecell}
\usepackage{stackengine}
\newcommand{\sket}[1]{{\ensuremath{\lvert#1\rangle}}}
\newcommand{\lket}[1]{{\ensuremath{\left\lvert#1\right\rangle}}}
\newcommand{\ket}[1]{\if@display\lket{#1}\else\sket{#1}\fi}
\begin{document}
\title{Spectral Attack on Continuous-Variable Quantum Key Distribution Systems}
\author{Chen Gong}
\affiliation{State Key Laboratory of Photonics and Communications, Institute for Quantum Sensing and Information Processing, Shanghai Jiao Tong University, Shanghai 200240, China}
\author{Mingxuan Guo}
\affiliation{State Key Laboratory of Photonics and Communications, Institute for Quantum Sensing and Information Processing, Shanghai Jiao Tong University, Shanghai 200240, China}
\author{Peng Huang}\email{huang.peng@sjtu.edu.cn}
\affiliation{State Key Laboratory of Photonics and Communications, Institute for Quantum Sensing and Information Processing, Shanghai Jiao Tong University, Shanghai 200240, China}
\affiliation{Shanghai Research Center for Quantum Sciences, Shanghai 201315, China}
\affiliation{Hefei National Laboratory, Hefei 230088, China}
\author{Tao Wang}
\affiliation{State Key Laboratory of Photonics and Communications, Institute for Quantum Sensing and Information Processing, Shanghai Jiao Tong University, Shanghai 200240, China}
\affiliation{Shanghai Research Center for Quantum Sciences, Shanghai 201315, China}
\affiliation{Hefei National Laboratory, Hefei 230088, China}
\author{Guihua Zeng}\email{ghzeng@sjtu.edu.cn}
\affiliation{State Key Laboratory of Photonics and Communications, Institute for Quantum Sensing and Information Processing, Shanghai Jiao Tong University, Shanghai 200240, China}
\affiliation{Shanghai Research Center for Quantum Sciences, Shanghai 201315, China}
\affiliation{Hefei National Laboratory, Hefei 230088, China}
\affiliation{Shanghai XunTai Quantech Co., Ltd, Shanghai, 200241, China}
	
\begin{abstract}
	Continuous-variable quantum key distribution (CVQKD) has attracted extensive attention due to its compatibility and low costs. However, bandwidth mismatch exists to varying degrees between the transmitter and receiver. This may prevent frequency components carrying modulation information from being fully perceived by the legitimate party. In this paper, we identify a practical security loophole caused by bandwidth mismatch and propose a corresponding spectral attack scheme. Different from previous approaches that exploit security loopholes to conceal the excess noise introduced by intercept-resend attacks, this scheme can directly obtain raw-key information without introducing additional disturbances. A proof-of-principle attack on a CVQKD system with filtering operation is constructed to verify the feasibility. Experimental results indicate that Eve can obtain enough information to render the system insecure if this practical security loophole is ignored. Based on the identified security loophole, corresponding defense strategies are proposed. This work helps bridge the gap between theoretical models and practical implementations, providing a reference for defense design in practical quantum communication systems.
\end{abstract}

\maketitle
\section{Introduction}\label{sec-1}
Quantum key distribution (QKD) is a key distribution technique that achieves information-theoretic security based on the principles of quantum mechanics~\cite{ekert1991quantum,lo1999unconditional,bennett2014quantum}.  Continuous-variable quantum key distribution (CVQKD) ~\cite{grosshans2002continuous,grosshans2003quantum,leverrier2009unconditional} is an important branch of QKD, which encodes information onto the quadrature components of the photonic optical field. It relies on mature coherent optical communication devices, and is highly compatible with classical coherent optical communication systems ~\cite{kumar2015coexistence,hajomer2025coexistence}. In recent years, substantial progress has been made in CVQKD in terms of theoretical studies ~\cite{leverrier2015composable,leverrier2017security,matsuura2021finite,lupo2022quantum}, experimental demonstrations~\cite{jouguet2013experimental,soh2015self,qi2015generating,huang2016long,zhang2020long,ren2021demonstration,xu2024robust,hajomer2024long}, post-processing techniques~\cite{cao2023rate,wang2023non,xing2025rate}, and network deployment~\cite{xu2023round,hajomer2024continuous,xu2025ofdm,qi2024experimental}, while analyses of its theoretical and practical security have also continued to evolve. Gaussian-modulated coherent states (GMCS) CVQKD is one of the most extensively studied protocols in CVQKD. Its theoretical security has been rigorously proven~\cite{grosshans2004continuous,navascues2006optimality,garcia2006unconditional,renner2009finetti,leverrier2013security}, and extensive achievements have been made in the analysis of the practical security of GMCS-CVQKD systems ~\cite{jain2022practical, kunz2015robust,ma2013wavelength,zheng2019practical}. Due to the imperfections of the practical devices, CVQKD systems inevitably suffer from security loopholes that can be exploited by Eve to conceal her attacks.

Currently, some typical attacks on practical CVQKD system are proposed, including calibration attacks \cite{jouguet2013preventing}, local-oscillator fluctuation attacks \cite{ma2013local}, and saturation attacks \cite{qin2016quantum}. Calibration attacks manipulate the local oscillator pulses to control shot noise, causing legitimate parties to misestimate the excess noise and thereby opening a security loophole for Eve. Local-oscillator fluctuation attacks conceal eavesdropping by misleading legitimate parties in their assessment of the system shot noise and excess noise. Saturation attacks exploit the limited linear operating range of the detector to make Bob’s detection results deviate from the ideal values, thereby hiding the noise introduced by intercept-resend attacks and stealing information about the secret key.

However, in previous studies on the practical security of CVQKD, security analyses have generally focused on signals within the effective detection bandwidth. At the same time, most existing attack schemes often struggle to directly obtain information related to the raw key without disrupting the legitimate communication process. Alternatively, they need to exploit practical security loopholes to cause the legitimate parties to underestimate the excess noise of the system, thereby concealing the additional excess noise introduced by intercept-resend attacks and ultimately accomplishing eavesdropping. However, previous studies have overlooked the fact that frequency components outside the effective detection bandwidth may still carry modulation information. This loophole may allow Eve to extract these out-of-band signal components and obtain raw key information without the need to conceal attack-induced disturbances.

In this paper, we investigate security issues caused by bandwidth mismatch between the transmitter and receiver in local local-oscillator (LLO) CVQKD system. From a frequency-domain perspective, we establish the relationship between the modulation information and the frequency components of the signal, showing that frequency components outside the effective bandwidth may lead to modulation information leakage. On this basis, a security loophole caused by bandwidth mismatch is revealed. Eve can extract out-of-band frequency components using a wavelength-division device, thereby obtaining raw key information. Different from previous attacks that exploit system imperfections to conceal the additional excess noise introduced by intercept-resend attacks, the eavesdropping scheme can directly extract raw key information without introduce extra disturbances. We construct a proof-of-principle attack on an LLO-CVQKD system with filtering operation for experimental demonstration. The results verify that eavesdroppers can obtain sufficient information from frequency components outside the effective detection bandwidth, thereby rendering the system insecure. Accordingly, corresponding defense strategies are proposed based on these findings, providing a reference for establishing a more comprehensive practical security architecture for CVQKD.

\section{GMCS-CVQKD system}\label{sec-2}
In this section, the specific protocol procedure and physical implementation of GMCS-CVQKD system are provided.

\begin{figure}[t]
	\centering
	\includegraphics[width=1\linewidth]{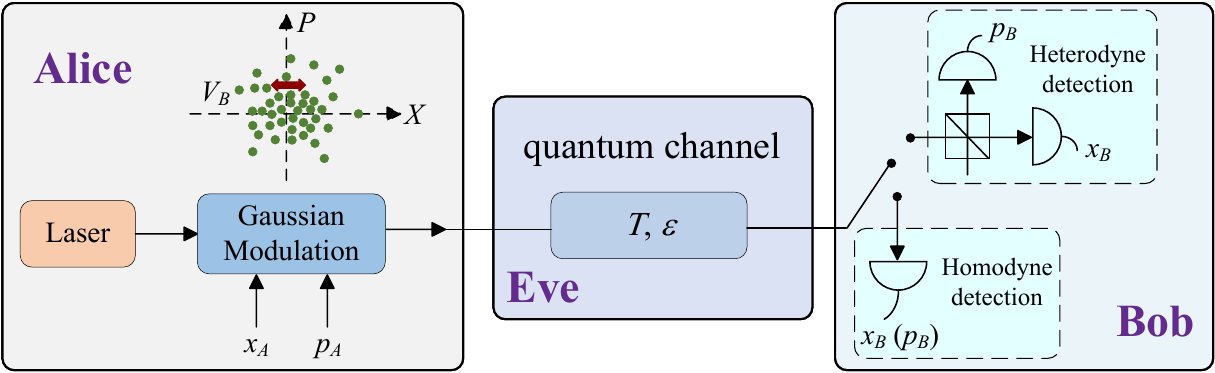}
	\caption{Prepare-and-measure model of the GMCS-CVQKD protocol.}\label{fig-1}
\end{figure}

The GMCS protocol, also known as the GG02 protocol, is one of the most representative protocols in CVQKD and has been widely applied in QKD systems. Its prepare-and-measure model is shown in Fig.~\ref{fig-1}. The specific procedure is as follows,

(a) Alice generates two independent and identically distributed Gaussian random sequences,
$\left\{ {{x_A}} \right\}$ and $\left\{ {{p_A}} \right\}$, with a mean of 0 and a variance of ${V_A}$. Subsequently, Alice prepares $N$ coherent states $\ket{x_A+ip_A}$ according to these two random sequences and sends them to Bob through the quantum channel with the channel loss $T$ and channel excess noise  $\varepsilon$.

(b) After receiving the quantum states, Bob may either randomly measure one quadrature component via homodyne detection or simultaneously measure both quadrature components using heterodyne detection.

(c) In the case of heterodyne detection, Alice retains all the data. In the case of homodyne detection, Bob publicly announces the basis selection information, based on which Alice discards part of the data.

(d) Alice and Bob randomly select a subset of the retained data for parameter estimation to calculate the channel loss and the channel excess noise, and determine the secret key rate based on these parameters.

(e) The communicating parties then perform data post-processing, including reverse reconciliation and privacy amplification, so that both parties ultimately obtain an identical bit string as the final secret key.

CVQKD systems can be classified into transmitted-local-oscillator (TLO) CVQKD systems and LLO-CVQKD systems according to whether the local oscillator is transmitted through the channel together with the quantum signal. In TLO-CVQKD systems, the local oscillator is vulnerable to manipulation by the eavesdropper, which poses security risks. Therefore, LLO-CVQKD is currently the mainstream approach. In LLO-CVQKD systems, the local oscillator light is not transmitted by Alice but is instead generated locally by Bob. This approach effectively reduces the risk of attacks associated with the local oscillator. However, to establish a global phase reference, Alice usually need to transmit reference pulses with quantum signal in LLO scheme.

\section{Spectral attack scheme}\label{sec-3}
\subsection{Relationship between modulation information and frequency components}\label{subsec3.1}
The transmitted signal is typically composed of two orthogonal Gaussian-modulated components. Let the two modulation variables at the $k{\rm{ - th}}$ symbol time be denoted as ${X_k}$ and ${P_k}$, respectively, with $g\left( t \right)$ denoting the pulse-shaping function and $T$ the symbol interval. Then, the in-phase and quadrature components of the transmitted signal can be expressed as,
\begin{align}
	{x_I}\left( t \right) =& \sum\limits_k {{X_k}g\left( {t - kT} \right)},\label{Eq.1}\\
	{x_Q}\left( t \right) =& \sum\limits_k {{P_k}g\left( {t - kT} \right)}.\label{Eq.2}
\end{align}
Therefore, its complex envelope is given by
\begin{align}
	s\left( t \right) =& {x_I}\left( t \right) + i{x_Q}\left( t \right)\notag \\
	=& \sum\limits_k {\left( {{X_k} + i{P_k}} \right)} g\left( {t - kT} \right).\label{Eq.3}
\end{align}
The frequency-domain expression of the transmitted signal is obtained by applying the Fourier transform to Equation (\ref{Eq.3}),
\begin{align}
	S(f) = G(f)\sum_k {({X_k}+i{P_k})e^{-i2\pi fkT}},\label{Eq.4}
\end{align}
where $G\left( f \right)$ is the Fourier transform of the pulse-shaping function $g\left(t\right)$.

As can be seen from the above formulas, the spectral component $S\left( f \right)$ of a transmitted signal at any frequency $f$ is determined by two terms. One part is the frequency-domain response $G\left( f \right)$ of the pulse-shaping function, while the other part is the weighted superposition term $\sum\limits_k {\left( {{X_k} + i{P_k}} \right){e^{ - j2\pi fkT}}}$ of the modulation sequence $\left\{ {{X_k},{P_k}} \right\}$. As long as $G\left( f \right) \ne 0$, the spectral component $S\left( f \right)$ varies with the transmitter-side modulation sequence. It can be seen that, within the effective bandwidth of the transmitted signal, the information is distributed across the entire effective bandwidth, with each non-zero frequency component carrying modulation information.

\subsection{Information leakage from out-of-band frequency components}\label{subsec3.2}
Based on the above analysis, it can be seen that the ideal spectrum of the CVQKD transmitter-side signal is determined jointly by the pulse-shaping function $g\left( t \right)$ and the modulation sequence, with its primary energy $k{E_0}\left( {k < 1} \right)$ (where ${E_0}$ denotes the signal energy) concentrated within the bandwidth ${B_0}$. However, in practical systems, the laser linewidth, phase noise, and the non-idealities in the transmitter and receiver-chain may cause the received signal spectrum to deviate from its ideal case. The actual received spectrum can be expressed as follows after considering the above effects,
\begin{align}
	{S_{\rm{act}}}\left( f \right) = {H_{\rm{trx}}}\left( f \right){H_{\rm{ch}}}\left( f \right)\tilde S\left( f \right),\label{Eq.5}
\end{align}
where ${H_{\rm{trx}}}\left( f \right)$ and ${H_{\rm{ch}}}\left( f \right)$ represent the transmitter and receiver -chain response and the channel response, respectively, while $\tilde S\left( f \right)$ denotes the actual spectrum after broadening caused by factors such as laser linewidth, frequency jitter, and phase noise.

If this spectral broadening effect is uniformly modeled as a spectral spreading kernel $L\left( f \right)$, then
\begin{align}
	\tilde S\left( f \right) = S\left( f \right) * L\left( f \right),\label{Eq.6}
\end{align}
where $*$ represents the convolution operation. By expanding Equation (\ref{Eq.6}), we get
\begin{align}
	\tilde S\left( f \right) = \int_{ - \infty }^{ + \infty } {S\left( v \right)} L\left( {f - v} \right)dv.\label{Eq.7}
\end{align}
The Equation (\ref{Eq.7}) indicates that as long as the spectral spreading kernel $L\left( f \right)$ has a nonzero width, the nonzero components of the ideal spectrum $S\left( f \right)$ within the bandwidth ${B_0}$ will spread into adjacent frequency regions. It means that non-ideal factors in the practical systems can lead to spectral broadening. Moreover, the expanded out-of-band frequency components are not completely independent of the original modulated signal. These components essentially originate from the action of the spectral spreading kernel $L\left( f \right)$ on the ideal modulation spectrum, they remain correlated with the information. On the other hand, the responses of the transmitter and receiver-chain and transmission channel further weight these broadened frequency components, thereby determining their actual retention in the system.
\begin{figure}[t]
	\centering
	\includegraphics[width=1\linewidth]{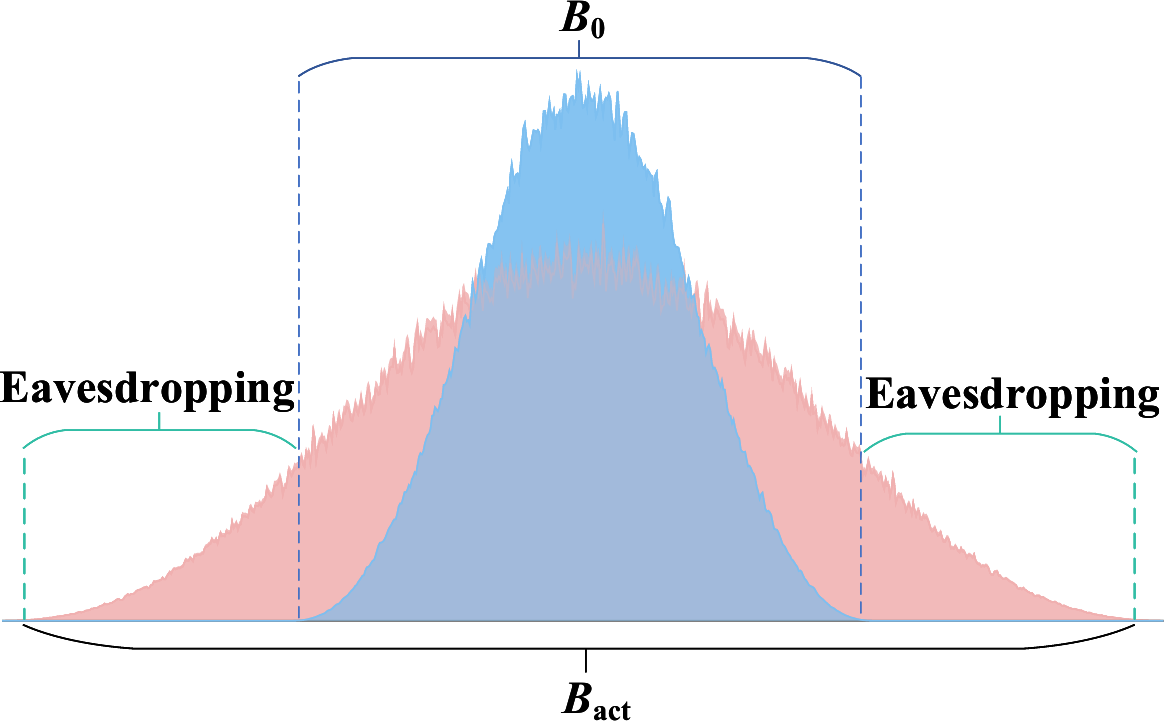}
	\caption{Schematic illustration of eavesdropping on out-of-band frequency components under bandwidth mismatch conditions, where blue denotes the transmitter signal and pink denotes the receiver detection signal.}\label{fig-2}
\end{figure}

In practical CVQKD systems, the effective bandwidth ${B_{\rm{rx}}}$ of the recevier and subsequent electronic circuitry is typically designed based on the bandwidth ${B_0}$ to ensure high-fidelity reception of in-band signals. However, the effective bandwidth of the actual signal may extend to ${B_{\rm{act}}} > {B_0}$ due to spectral spreading. When ${B_{\rm{act}}} > {B_{\rm{rx}}}$, some out-of-band frequency components located outside ${B_0}$ may be difficult for Bob to detect effectively or estimate accurately. In this scenario, Eve can selectively extract frequency components satisfying $f \notin {B_{\rm{rx}}}$ using a wavelength splitter device, and obtain additional information by exploiting the spectral components associated with the modulation infromation. Since the detection process of Bob is confined to its effective bandwidth ${B_{\rm{rx}}}$, Eve can perform covert attacks while minimizing disturbance to the normal detection results of the system. The specific implementation is shown in Fig.~\ref{fig-2}.

\subsection{Eavesdropping feasibility}\label{subsec3.3}
To verify the feasibility of the proposed attack scheme, numerical simulations are conducted in this section to analyze the attack effects under various bandwidth mismatch conditions.

Ideally, we can assume that the wavelength splitter device has no insertion loss, so the spectral attack will not disturb legitimate communication process. In this case, any amount of information obtained by Eve may potentially render the CVQKD system insecure. However, in practice, the wavelength splitter device inevitably induces a certain insertion loss, which can be reflected in the parameter estimation performed by the legitimate patries. The insertion loss of the wavelength splitter device will affect the difficulty of implementing the attack to a certain extent. Therefore, ie is necessary to calculate the amount of information that Eve needs to eavesdrop under different insertion loss conditions to render the system insecure. Thus, the additional amount of information ${I_{{\rm{ext}}}}$ that Eve may to obtain under the spectral attack to render the system insecure is derived. The estimated channel loss serves as the insertion loss ${T_{{\mathop{\rm int}} }}$ of the wavelength splitter from the perspective of the legitimate parties. Furthermore, the amount of information eavesdropped by Eve as estimated by the legitimate parties is ${\chi _{{\rm{bE}}}}\left( {T = {T_{{\mathop{\rm int}} }},\xi  = 0} \right)$. Meanwhile, the information actually obtained by Eve in the spectral attack is defined as ${I_{{\rm{SA}}}}$. Thus, the total amount of information actually obtained by Eve must exceed the amount of information estimated by the legitimate parties for successful eavesdropping, i.e.,
\begin{align}
	{I_{{\rm{SA}}}} \ge {\chi _{{\rm{bE}}}}\left( {T = {T_{{\mathop{\rm int}} }},\xi  = 0} \right).\label{Eq.8}
\end{align}
Hence, ${I_{{\rm{ext}}}} = {\chi _{{\rm{bE}}}}\left( {T = {T_{{\mathop{\rm int}} }},\xi  = 0} \right)$ and we can give out the relationship between ${\rm{SN}}{{\rm{R}}_{be}}$ (signal-to-noise ratio of the information for Eve and Bob) and ${I_{{\rm{SA}}}}$,
\begin{align}
	{\rm{SN}}{{\rm{R}}_{be}} = {2^{{I_{{\rm{SA}}}}}} - 1.\label{Eq.9}
\end{align}
Based on this, the minimum signal-to-noise ratio required by Eve in the spectral attack can be easily obtained. Next, the expressions for the information obtained by the receiver and the eavesdropper in the spectral attack are further derived,
\begin{align}
	{X_B} =& \sqrt {0.5{\eta _B}{T_{{\mathop{\rm int}} }}V_A^B} \hat X + N_X^B, \label{Eq.10}\\
	{P_B} =& \sqrt {0.5{\eta _B}{T_{{\mathop{\rm int}} }}V_A^B} \hat P + N_X^B,\label{Eq.11}\\
	{X_E} =& \sqrt {0.5{\eta _E}{T_{{\mathop{\rm int}} }}V_A^E} \hat X + N_X^E, \label{Eq.12}\\
	{P_E} =& \sqrt {0.5{\eta _E}{T_{{\mathop{\rm int}} }}V_A^E} \hat P + N_P^E,\label{Eq.13}\\
	Var\left( {N_X^B} \right) =& Var\left( {N_P^B} \right) = 1 + v_{{\rm{el}}}^B, \label{Eq.14}\\
	Var\left( {N_X^E} \right) =& Var\left( {N_P^E} \right) = 1 + v_{{\rm{el}}}^E,\label{Eq.15}
\end{align}
where ${\eta _B},N_X^B,N_P^B$ denote the detection efficiency and the noise on the $\hat X$ and $\hat P$ quadratures at the receiver, respectively, while ${\eta _E},N_X^E,N_P^E$ denote these at the eavesdropper. $\hat X$ and $\hat P$ represent the normalized modulation information. $v_{{\rm{el}}}^B$ and $v_{{\rm{el}}}^E$ represent the electronic noise at the receiver and the eavesdropper, respectively. $V_A^B$ indicates the modulation variance normalized to Alice site after Bob applies a low-pass filtering to the received signal and $V_A^E$ represents the modulation variance normalized to Alice site after Eve applies high-pass filtering to the received signal. The expressions for the normalized $\hat X$ and $\hat P$ quadratures at both the eavesdropper and the receiver can be determined,
\begin{align}
	X_B' =& \hat X + \frac{{N_X^B}}{{\sqrt {0.5{\eta _B}{T_{{\mathop{\rm int}} }}V_A^B} }},\label{Eq.16}\\
 X_E'=&X_B'-\frac{N_X^B}{\sqrt{0.5\eta_BT_{\mathrm{int}}V_A^B}}+\frac{N_X^E}{\sqrt{0.5\eta_ET_{\mathrm{int}}V_A^E}},\label{Eq.17}\\
	P_B'=&\hat P+\frac{N_P^B}{\sqrt{0.5\eta_BT_{\mathrm{int}}V_A^B}},\label{Eq.18}\\ P_E'=&P_B'-\frac{N_P^B}{\sqrt{0.5\eta_BT_{\mathrm{int}}V_A^B}}+\frac{N_P^E}{\sqrt{0.5\eta_ET_{\mathrm{int}}V_A^E}}.\label{Eq.19}
\end{align}

Assuming that the signal-to-noise ratio of the $\hat X$ and $\hat P$ quadratures are identical, the relationship between $V_A^E$ and SNR is given by
\begin{align}
	{\rm{SN}}{{\rm{R}}_{be}}=& \frac{{1}}{{Var(\frac{N_X^B}{\sqrt {0.5{\eta _B}{T_{{\mathop{\rm int}} }}V_A^B}}  - \frac{N_X^E}{\sqrt {0.5{\eta _E}{T_{{\mathop{\rm int}} }}V_A^E}} )}} \notag\\
	=& \frac{1}{{\frac{{1 + v_{el}^B}}{{0.5{\eta _B}{T_{{\mathop{\rm int}} }}V_A^B}} + \frac{{1 + v_{el}^E}}{{0.5{\eta _E}{T_{{\mathop{\rm int}} }}V_A^E}}}}.\label{Eq.20}
\end{align}
Accordingly, the required minimum $V_A^E$ for the spectral attack can also be determined.
\begin{figure*}[t]
	\centering
	\includegraphics[width=\linewidth]{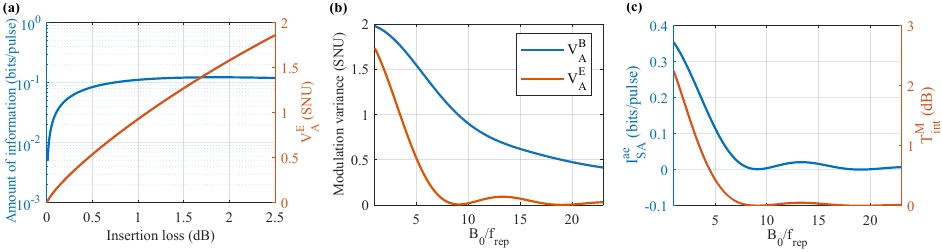}
	\caption{Simulation results. (a) Required eavesdropped information and corresponding minimum required $V_A^E$ for rendering the system insecure under different insertion losses, where $V_A^B$ is 2 shot noise units (SNU); (b) The relationship of $V_A^B$ and $V_A^E$ with the normalized bandwidth $B_0/f_{\mathrm{rep}}$; (c) The relationship of $I_{\rm{SA}}^{ae}$ and the maximum allowable insertion loss $T_{{\mathop{\rm int}} }^M$ with the normalized bandwidth $B_0/f_{\mathrm{rep}}$. The detection efficiency is 0.48, the electrical noise is 0.5 SNU, the peak value of the transmitter-side signal spectrum is $\sqrt{2}$, the repetition frequency is 500 MHz, the range of the bandpass filtering at Eve is $\left[ {{B_0}/2,{B_0}/2 + {f_{{\rm{rep}}}}} \right]$, and the transmitter-side pulse is a rectangular pulse.}\label{fig-3}
\end{figure*}

Fig.~\ref{fig-3}(a) presents the additional amount of eavesdropped information $I_{\mathrm{ext}}$ required by Eve to render the system insecure via the spectral attack under different insertion losses of the wavelength splitter, as well as the corresponding minimum required $V_A^E$. It can be seen that both $I_{\mathrm{ext}}$ and $V_A^E$ show an increasing trend as the insertion loss increases. This indicates the greater the insertion loss, the more information Eve must obtain to offset the increased channel loss caused by the loss and keep the system insecure. Therefore, a larger $V_A^E$ is required for the out-of-band signal obtained by Eve.

Subsequently, we further analyze the eavesdropping ability of spectral attack under rectangular pulse shaping with different receiver bandwidths. The frequency-domain expression of a rectangular pulse is,
\begin{align}
	{X_{{\rm{Rec}}}}\left( f \right) = A\tau  \cdot {\rm{sinc}}\left( {f\tau } \right),\label{Eq.21}
\end{align}
where $\tau  = D/{f_{{\rm{rep}}}}$ denotes the pulse width, $D$ denotes the duty cycle, and $A$ denotes the intensity coefficient. Since the spectrum of shot noise is flat, $V_A^E$ and $V_A^B$ are given as follows,

\begin{align}
	V_A^E =& \frac{{\int_{ - \frac{B_0}{2}}^{ - \frac{B_E}{2}} {{{\left| {{X_{{\rm{Rec}}}}\left( f \right)} \right|}^2}df + \int_{\frac{B_0}2}^{\frac{B_E}2} {{{\left| {{X_{{\rm{Rec}}}}\left( f \right)} \right|}^2}df} } }}{{\int_{ - \frac{B_0}2}^{ - \frac{B_E}2} {1 \cdot df + \int_{\frac{B_0}2}^{\frac{B_E}2} {1 \cdot df} } }}\notag\\
	=& \frac{{2 \cdot \int_{\frac{B_0}2}^{\frac{B_E}2} {{{\left| {{X_{{\rm{Rec}}}}\left( f \right)} \right|}^2}df} }}{{{B_E} - {B_0}}},\label{Eq.22}\\
	V_A^B =& \frac{{\int_{ - \frac{B_0}2}^{\frac{B_0}2} {{{\left| {{X_{{\rm{Rec}}}}\left( f \right)} \right|}^2}df} }}{{\int_{ - \frac{B_0}2}^{\frac{B_0}2} {1 \cdot df} }} = \frac{{\int_{ - \frac{B_0}2}^{\frac{B_0}2} {{{\left| {{X_{{\rm{Rec}}}}\left( f \right)} \right|}^2}df} }}{{{B_0}}}.\label{Eq.23}
\end{align}

According to the above equations, $V_A^E$, $V_A^B$, and $I_{\mathrm{SA}}^{ae}$ under different bandwidth conditions can be calculated, where $I_{\mathrm{SA}}^{ae}$ refers to the mutual information between the information obtained by spectral attack and the modulated information. And the maximum insertion loss of the wavelength splitter that Eve can tolerate when successfully implementing the spectral attack $T_{\mathrm{int}}^{\mathrm{M}}$ can also be calculated. The relationships of the above metrics with the normalized bandwidth ${B_0}/{f_{{\rm{rep}}}}$ are presented in Figures~\ref{fig-3}(b) and (c), respectively. As shown in Fig.~\ref{fig-3}(b), both $V_A^B$ and $V_A^E$ exhibit a decreasing trend, while $V_A^E$ shows a slowly fluctuating behavior. The frequency spectrum envelope of the transmitted signal for rectangular pulse shaping follows a $\mathrm{sinc}^2$ distribution, and thus its spectral energy gradually decreases with increasing frequency. The low-pass filtering window corresponding to Bob always covers the low-frequency main-lobe region of the signal, so $V_A^B$ decreases as the normalized bandwidth increases. In contrast, the high-pass filtering window corresponding to Eve is located in the out-of-band frequency range, and these spectral components fall within the side lobe region of the rectangular-pulse spectrum. Since the side lobes energy is relatively weak and decays continuously with increasing frequency,  $V_A^E$ remains significantly lower than $V_A^B$ overall and further decreases as $B_0/f_{\mathrm{rep}}$ increases.

Fig.~\ref{fig-3}(c) shows the relationship of $I_{\mathrm{SA}}^{ae}$ and $T_{\mathrm{int}}^{\mathrm{M}}$ with respect to $B_0/f_{\mathrm{rep}}$, respectively. It can be observed that $I_{\mathrm{SA}}^{ae}$, generally decreases as $B_0/f_{\mathrm{rep}}$ increases. This indicates that, as the effective bandwidth of the receiver gradually increases, the out-of-band eavesdropping window continuously shifts toward higher-frequency regions with relatively weak spectral energy. Consequently, the amount of valid information she can extract gradually decreases. $T_{{\mathop{\rm int}} }^{\mathrm{M}}$ exhibits an overall decreasing trend as $B_0/f_{\mathrm{rep}}$ increases. This suggests that increasing the normalized bandwidth of the receiver significantly reduces Eve's tolerance to the insertion loss of the wavelength splitter. A smaller $B_0/f_{\mathrm{rep}}$ allows a greater maximum insertion loss, indicating that the out-of-band frequency components still contain strong modulation-related information. Thus, Eve may still successfully implement the attack even in the presence of a certain insertion loss.

\section{Experimental verification}\label{sec-4}
\subsection{Experimental scheme}\label{subsec4.1}
Currently, common local oscillator schemes include polarization-division-multiplexed (PDM) LLO-CVQKD and frequency-division-multiplexed (FDM) LLO-CVQKD, in which the pilot and quantum signals are transmitted simultaneously through plarization-division multiplexing and frequency-division multiplexing, respectively. The specific implementation of the proposed attack scheme in these two types of systems is described in detail below.
\begin{figure*}[t]
	\centering
	\includegraphics[width=0.9\linewidth]{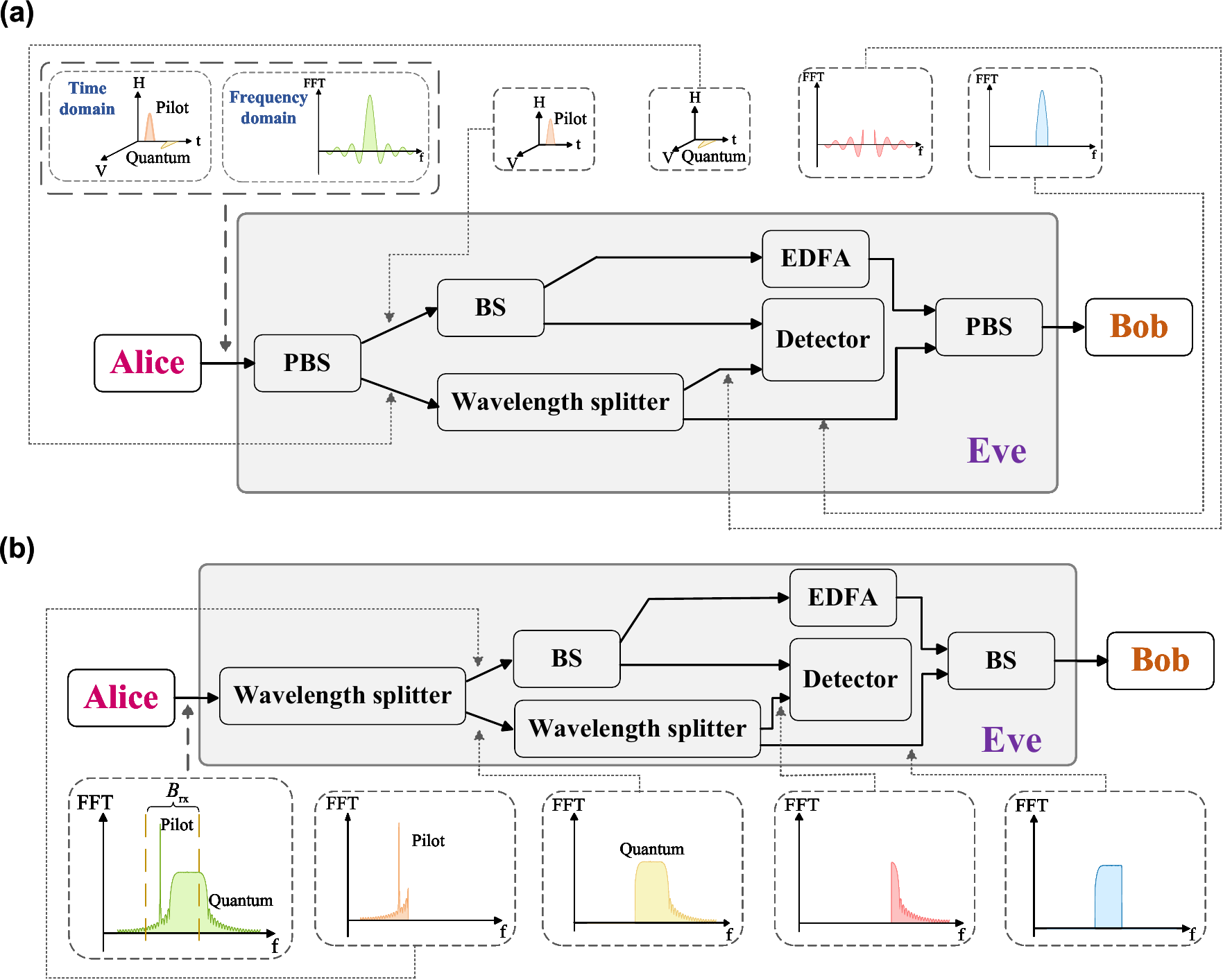}
	\caption{Schematic diagram of the spectral attack based on bandwidth mismatch. (a) Attack scheme for a PDM-LLO-CVQKD system; (b) Attack scheme for a FDM-LLO-CVQKD system.}\label{fig-4}
\end{figure*}

As shown in Fig.~\ref{fig-4}(a), in the PDM-LLO-CVQKD system, Eve employs a polarization beam splitter (PBS) to separate the polarization-multiplexed light emitted by Alice, thereby directing the pilot signal and the quantum signal into different paths. After being further split by a beam splitter (BS), one output of the pilot signal in the upper path is sent to a detector to obtain phase reference information. The other path is sent to an erbium-doped fiber amplifier (EDFA) for amplification to improve the signal-to-noise ratio of the pilot used by the legitimate parties, thereby preventing the eavesdropping operation from introducing additional phase noise to the legitimate parties. The quantum signal in the lower path is then fed into a wavelength splitter, which separates the frequency components within and outside the effective bandwidth of Bob. The frequency components within the effective bandwidth are polarization-combined with the pilot signal amplified by the EDFA via a PBS to achieve polarization multiplexing, and then sent to Bob to prevent the legitimate parties from perceiving the eavesdropping operation. The frequency components outside the effective bandwidth are sent to a heterodyne detector to obtain the $X$ and $P$ quadratures. Afterward, based on the detection results of the pilot signals split by the BS, post-processing operations such as frequency offset recovery and phase recovery are performed to recover the modulation information carried by these components.
\begin{figure*}[t]
	\centering
	\includegraphics[width=\linewidth]{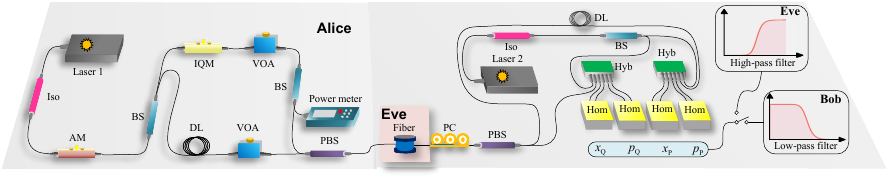}
	\caption{Optical schematic for validating the feasibility of the proposed spectral attack scheme. Iso optical isolator, AM amplitude modulator, BS beam splitter, IQM in-phase and quadrature modulator, VOA variable attenuator, DL delay line, PBS polarization beam splitter, COL Collmator, PC Polarization Controller, Hyb $90^\circ$ optical hybrid, Hom homodyne detector.}\label{fig-5}
\end{figure*}

Fig.~\ref{fig-4}(b) illustrates the specific implementation of the attack scheme in the FDM-LLO-CVQKD system. To begin with, Eve separates the frequency-multiplexed light emitted by Alice using a first-stage wavelength splitter, directing the pilot signal and the quantum signal into different paths. Similar to the PDM scheme, the pilot branch is split by a BS, and one output is sent to an EDFA for amplification to improve the signal-to-noise ratio of the pilot utilized by the legitimate parties. The other output is sent to a detector to acquire phase reference information. The quantum signal branch is further separated by a second-stage wavelength splitter into frequency components within and outside the effective bandwidth for Bob. The frequency components within the effective bandwidth are combined with the pilot signal amplified by the EDFA via a BS to achieve frequency multiplexing, and then sent to Bob. The subsequent processing of frequency components outside the effective bandwidth is consistent with the PDM-LLO-CVQKD scheme.

\subsection{Experimental implementation}\label{subsec4.2}
To demonstrate the effectiveness of the spectral attack scheme, a LLO GMCS-CVQKD system is established in this work, with its experimental setup shown in Fig.~\ref{fig-5}. This scheme multiplexes the pilot signal and the quantum signal in both the time domain and the polarization domain \cite{wang2018pilot,wu2021passive}.

At the transmitter, a narrow-linewidth continuous-wave laser 1 (NKT Koheras BASIK E15) with a linewidth of 100 Hz and a center wavelength of 1550.12 nm is employed to generate the optical carrier. The optical carrier passes through an optical isolator (Iso) to prevent reflected light from re-entering the laser, and then enters an amplitude modulator (AM; KY-AM-15-BW-PP-40dB), where it is subsequently modulated to form a pulsed optical signal. The AM is driven by an analog electrical signal from an arbitrary waveform generator (AWG; Tektronix, AWG5200) and biased using a stable direct-current (DC) power supply. The continuous light generated by the laser is cut by the AM into an optical pulse sequence with a duty cycle of 20\% and a repetition rate of 50 MHz. Subsequently, the optical pulse sequence is split into two paths using a BS, with one path serving as the signal light and the other as the pilot light. The signal light is sent into an in-phase and quadrature modulator (IQM) to achieve Gaussian modulation. The IQM is controlled by analog electrical signals generated by the AWG based on two sets of independent Gaussian random numbers. The intensity of the modulated quantum signal light is adjusted using a VOA to control its modulation variance. To monitor the intensity of the transmitted signal, a BS is connected to the output of the VOA to split off a small portion of the light for measuring the intensity of the quantum signal.

In the pilot optical path, a DL is employed to delay the pilot pulse, temporally offset from the quantum signal pulse to achieve time-division-multiplexing. Next, the intensity of the pilot light is adjusted using a VOA. Finally, the quantum signal and the pilot signal are polarization-combined using a polarization beam splitter (PBS) with a polarization isolation of 30 dB to achieve polarization-multiplexing, and then jointly transmitted through the fiber channel with 8 km and 25 km.
\begin{figure*}[t]
	\centering
	\includegraphics[width=\linewidth]{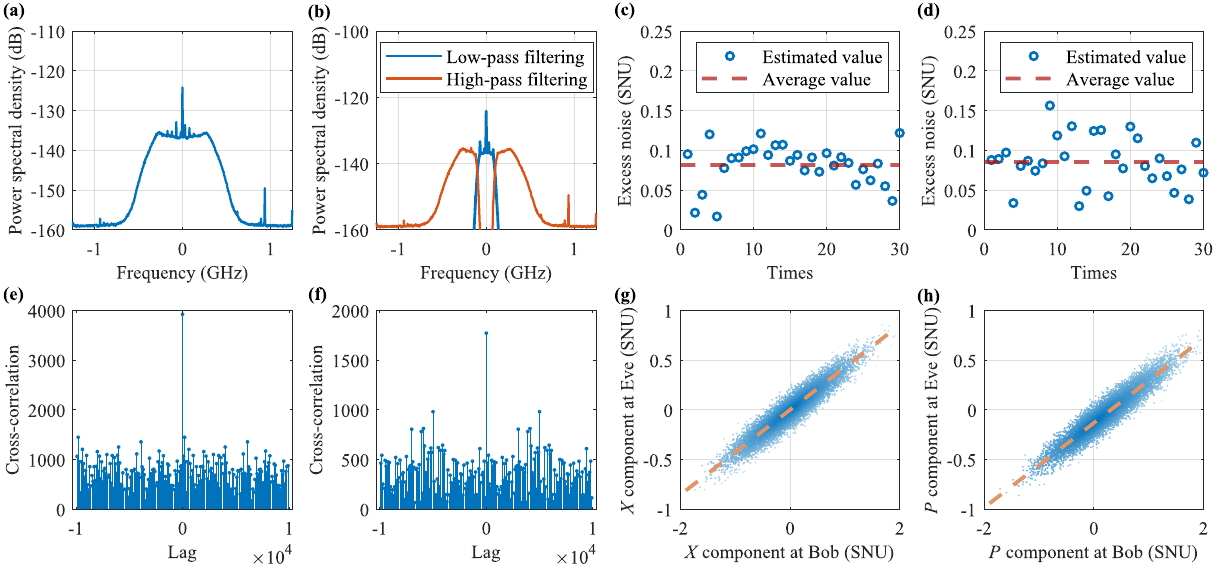}
	\caption{Experimental results. (a) Spectrum of the original received signal; (b) Spectrum of the signal after low-pass and high-pass filtering; Estimated excess noise at a transmission distance of (c) 8 km and (d) 25 km; Cross-correlation results between Eve’s eavesdropped information and Bob’s detected information at (e) 8 km and (f) 25 km; Relationship between the (g) $X$ and (h) $P$ quadrature detection results of Bob and Eve at a transmission distance of 8 km.}\label{fig-6}
\end{figure*}

At the receiver, the orthogonally polarization-multiplexed light received by Bob is adjusted by a PC, such that the quantum signal light is aligned with the $H$ polarization and the pilot light is aligned with the $V$ polarization. Subsequently, the light adjusted by the PC is polarization-demultiplexed using a PBS. The two optical beams separated by the PBS correspond to the pilot pulse and the quantum signal pulse, respectively. Since the quantum signal pulse and the pilot pulse are time-division-multiplexed, a DL is further introduced along the quantum signal path to realign the temporally offset quantum signal pulse with the pilot pulse, thereby achieving time-domain-demultiplexing. The demultiplexed quantum signal pulse and the pilot pulse are respectively sent to two optical hybrids (Hyb) and subsequently injected into the balanced homodyne detector (Hom; Thorlabs PDB435C-AC). Meanwhile, another narrow-linewidth continuous-wave laser 2 (NKT Koheras BASIK X15) with a linewidth of 100 Hz generates an optical carrier with a center wavelength of 1550.12 nm as the LO. To ensure the stability of the coherence detection, the wavelengths of Laser 1 and Laser 2 are precisely calibrated in advance using a spectrometer. The LO light is then split into two beams by a 50:50 BS and injected into two Hom detectors via two Hybs.

The coherent detection results are sampled by an 8-bit oscilloscope (LeCroy, WaveMaster 813Zi-B) with a sampling rate of 2.5 Gsamples/s, which corresponds to 50-fold oversampling of the 50 MHz signal. The system detection efficiency $\eta$ and electronic noise $v_{\rm{el}}$ are 0.48 and 0.4955 SNU, respectively. The electronic noise is calibrated using data collected without injecting the LO into the coherent detector. The single-frame data length is set to $10^6$ in the experiment.

After completing coherent detection and sampling, digital processing is performed on the detection results to simulate how Bob and Eve utilize different frequency-domain components of the received signal. Specifically, low-pass filtering is applied to the sampled data to extract the in-band signal components of Bob, thereby enabling parameter estimation and information recovery by the legitimate users. At the same time, high-pass filtering is applied to extract out-of-band frequency components, which are then regarded as the eavesdropping signal available to Eve. Thus, Bob’s in-band detection results and Eve’s out-of-band detection results can be obtained from the experimental data, thereby experimentally validating the feasibility of the proposed spectral attack based on bandwidth mismatch.
\begin{figure*}[t]
	\centering
	\includegraphics[width=\linewidth]{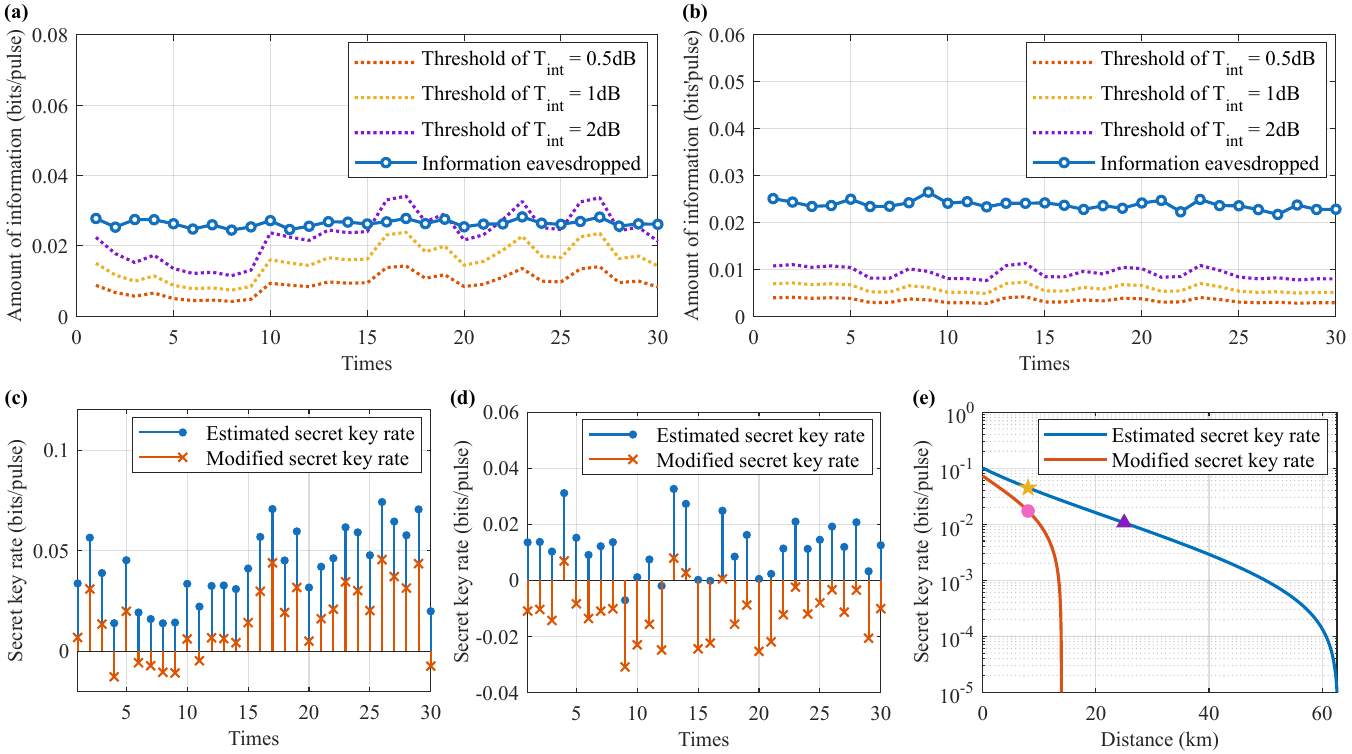}
	\caption{Attacking results with cutoff frequency 100 MHz. Comparison of the information obtained by Eve via spectral attack with the minimum information required to render the system insecure at (a) 8 km and (b) 25km; Secret key rates obtained from parameter estimation and modified secret key rates after considering spectral attack at (c) 8 km and (d) 25 km; (e) Secret key rates obtained from parameter estimation and modified secret key rates as a function of distance. The secret key rate curves are simulated based on average values for the estimated experimental parameters of each frame. The yellow pentagon, red circle and purple triangle represent experimental points at 8 km, 8km and 25 km. The modified secret key rate at 25 km is negative and is not plotted.}\label{fig-7}
\end{figure*}

\subsection{Experimental results}\label{subsec4.3}
After completing coherent detection and data sampling, the received data are processed using offline digital filtering to simulate the utilization of different frequency-domain components by Eve and Bob. Considering the advantages of FIR filters, such as linear phase response, simple implementation, and the convenience of precisely controlling the cutoff frequency, FIR digital filters designed using the window function method are employed to process the sampled data, with a Hamming window used to construct the finite impulse response filter. To minimize the phase distortion introduced by filtering, zero-phase filtering is adopted in the subsequent offline data processing. Specifically, a low-pass filter is utilized in the Bob branch to extract the low-frequency components of the received signal within the cutoff bandwidth, whereas a high-pass filter is employed in the Eve branch to extract the high-frequency components above the cutoff bandwidth. The two types of filters are designed with the same cutoff frequency $f_c$, thereby dividing the same received spectrum into low-frequency and high-frequency parts, with the cutoff frequency set to 100 MHz. After the filtering operation, we conduct the post-processing of the sampled data to compensate for the frequency offset and phase drift. 

Fig.~\ref{fig-6}(a) and (b) show the filtering results of the received signal in the frequency domain, where they depict the spectrum of the heterodyne detection signal after low-pass and high-pass filtering, respectively. The in-band frequency components near the center can be extracted via low-pass filtering, whereas the out-of-band frequency components located on both sides can be separated via high-pass filtering. Under bandwidth mismatch, the legitimate receiver primarily relies on the in-band signal, while the out-of-band frequency components are extracted by Eve to implement the spectral attack.

Channel parameter is one of the key parameters characterizing the performance and security of CVQKD systems, and its magnitude directly affects the final extractable secret key rate. After completing the post-processing, system parameters are estimated based on the correlation between the modulation data and the data received by Bob. A total of 30 data frames are collected, with a frame length of ${10^6}$, the channel loss and excess noise are further calculated. We obtain the channel loss for 8 km channel and 25 km channel are 2.86 dB and 5.47 dB, respectively. It is noted that the loss of the channel includes the loss of optical fiber and the insertion loss of interfaces. Fig.~\ref{fig-6}(c) and (d) display the distributions of the excess noise for each data frame at transmission distances of 8 km and 25 km, respectively. It can be noted that the excess noise at both transmission distances fluctuates around their respective mean values, but overall remains within a limited range. The statistical results show that the average excess noise values at 8 km and 25 km are 0.0817 SNU and 0.0855 SNU, respectively.

Fig.~\ref{fig-6}(e) and (f) present the cross-correlation results between the out-of-band information obtained by Eve and the in-band information detected by Bob at transmission distances of 8 km and 25 km. It can be observed that the cross-correlation curves at both distances exhibit a distinct main peak at the zero-lag position. This suggests that a significant statistical correlation exists between the out-of-band frequency components extracted by Eve and the in-band signal recovered by Bob. This illustrates that bandwidth mismatches between the transmitter and receiver may cause some frequency components to be insufficiently utilized by the legitimate receiver, making them available for extraction and recovery by attackers.

Fig.~\ref{fig-6}(g) and (h) depict the relationships between Bob’s and Eve’s detection results for the $X$ and $P$ quadratures. It can be noted that, for both the $X$ and $P$ quadratures, the scatter points cluster in a band-like pattern. The scattered points in the figure exhibit a certain degree of dispersion around the ideal fitted line, indicating that the information recovered by Eve is affected by factors such as channel noise and shot noise. Nevertheless, this dispersion does not undermine the overall linear relationship between the two. The information obtained by Eve is not random noise, but rather valid information that has a clear correspondence with modulation data. The $X$ and $P$ quadratures recovered from the out-of-band frequency components extracted by Eve retain a linear relationship with Bob's detection results.

Fig.~\ref{fig-7}(a) and (b) provide the variation in the amount of information obtained by Eve through the spectral attack at transmission distances of 8 km and 25 km. The blue curve represents the amount of information actually obtained by Eve in each experiment, while the three dashed lines represent the minimum information thresholds required to render the system insecure when the insertion losses of the wavelength splitter are 0.5 dB, 1 dB, and 2 dB, respectively. Since the practical CVQKD system is connected to fiber links and suffers from device imperfections, untrusted channel loss and channel noise are inevitably present. The amount of eavesdropping information associated with these untrusted parameters should be considered. Accordingly, Equation (\ref{Eq.8}) is modified as follows,
\begin{align}
	&{I_{{\rm{SA}}}} + {\chi _{{\rm{bE}}}}\left( {{V_A} = {T_{{\mathop{\rm int}} }}V_A^B,T = \overline{T_{\mathrm{c}}},\xi  = \overline \xi  } \right) \notag \\
	\ge& {\chi _{{\rm{bE}}}}\left( {{V_A} = V_A^B,T = {T_{{\mathop{\rm int}} }}\overline{T_{\mathrm{c}}},\xi  = \overline \xi  /{T_{{\mathop{\rm int}} }}} \right),\label{Eq.24}
\end{align}
where $\overline \xi  $ denotes the average value of the estimated excess noise and $\overline{T_{\mathrm{c}}}$ denotes the average value of the estimated channel loss. Therefore, the minimum information threshold required to render the system insecure in this case is ${I_{{\rm{ext}}}} = {\chi _{{\rm{bE}}}}\left( {{V_A} = V_A^B,T = {T_{{\mathop{\rm int}} }}\overline{T_{\mathrm{c}}},\xi  = \overline \xi  /{T_{{\mathop{\rm int}} }}} \right) - {\chi _{{\rm{bE}}}}\left( {{V_A} = {T_{{\mathop{\rm int}} }}V_A^B,T = \overline{T_{\mathrm{c}}},\xi  = \overline \xi  } \right)$. At a transmission distance of 8 km, the amount of information actually acquired by Eve is higher than the thresholds corresponding to $T_{\mathrm{int}} = 0.5$dB and $T_{\mathrm{int}} = 1$dB for all experimental points, and is also higher than the threshold corresponding to $T_{\mathrm{int}} = 2$dB in most cases. This indicates that even with a certain amount of insertion loss in the wavelength splitter, Eve can still stably extract sufficient information from the out-of-band frequency components, thereby placing the system in an insecure state. Furthermore, the blue curve remains significantly above the corresponding information thresholds for all three insertion loss values at a transmission distance of 25 km. This is because as the transmission distance increases, the tolerance of the system to additional information leakage decreases, allowing the spectral attack to make the system insecure with just a smaller amount of obtained information.

Fig.~\ref{fig-7}(c) and (d) provide the system secret key rates at transmission distances of 8 km and 25 km. The blue curves represent the original secret key rates obtained by the legitimate parties using conventional parameter estimations, while the orange curves represent the actual secret key rates modified by considering the impact of the spectral attack. At a transmission distance of 8 km, the modified actual secret key rate is generally lower than the original estimated secret key rate, indicating that the spectral attack reduces the amount of secret key. The reason is that Eve can extract additional information from the out-of-band frequency components, which this information has not fully accounted for by legitimate parties in conventional parameter estimation, thereby causing the original secret key rate to be overestimated. At a transmission distance of 25 km, the spectral attack has a more significant impact on the secret key rate. This implies that increasing the transmission distance makes the secret key rate estimated by the legitimate parties more prone to overestimating the actual security of the system. Fig.~\ref{fig-7}(e) visually illustrates this trend. The original estimated secret key rate decreases gradually as the channel loss increases, whereas the actual secret key rate decreases more rapidly after considering the spectral attack. These results demonstrate that out-of-band information leakage caused by bandwidth mismatch can significantly affect the final security boundary of the system. Therefore, it must be fully taken into account in the security analysis and parameter estimation of practical CVQKD systems.

To defend against the spectral attack based on bandwidth mismatch proposed in this paper, countermeasures can be implemented in terms of bandwidth design and secret key rate correction. First, at the system design stage, factors such as the repetition frequency, pulse shaping methods, frequency response of the transmission link, and laser linewidth should be comprehensively considered to reasonably design the effective detection bandwidth of the receiver. This approach minimizes the bandwidth mismatch between the transmitter and receiver as much as possible, thereby reducing the possibility of eavesdropping by attackers exploiting out-of-band frequency components. Subsequently, at the security estiamtion stage, the signal spectrum outside the receiver’s bandwidth can be estimated by analyzing relevant parameters. The modulation variance of the signal received by the eavesdropper under the spectral attack scenario can be obtained, and the maximum amount of information that the attacker may acquire under this condition can be calculated accordingly. On this basis, the additional leaked information is incorporated into the secret key rate calculation to modify the traditional secret key rate formula, thereby obtaining a revised secret key rate that more accurately reflects the security boundary of practical systems and provides a reference for subsequent privacy amplification.

\section{Conclusion}\label{sec-5}
In this study, we investigate the impact of bandwidth mismatch between the transmitter and receiver on the security of LLO-CVQKD systems, and propose a spectral attack scheme using out-of-band frequency components. Different from previous approaches that exploit practical security loopholes to conceal the excess noise introduced by intercept-resend attacks, this scheme can directly obtain raw-key information without introducing additional disturbances theoretically. The eavesdropping capability of the attacker is evaluated and simulated under various bandwidth mismatch conditions. Furthermore, we construst a proof-of-principle attack on an LLO-GMCS-CVQKD system with filtering operation to validate the proposed scheme. Experimental results indicate that the information obtained by Eve from the out-of-band frequency components is significantly correlated with the in-band information detected by Bob, and exhibits a linear correspondence across the two quadratures. Furthermore, when the impact of the spectral attack is considered, the actual secret key rate of the system is significantly lower than the original secret key rate estimated using conventional parameter estimation. To address the aforementioned security risks, corresponding defense strategies are designed in terms of bandwidth design and secret key rate correction, which may provide a more stable and efficient solution for the security analysis and defense design of large-scale CVQKD systems. 

\begin{acknowledgments}
	This work was supported by Quantum Science and Technology-National Science and Technology Major Project (Grant No. 2021ZD0300703), Shanghai Municipal Science and Technology Major Project (2019SHZDZX01).
\end{acknowledgments}

\bibliography{sacvqkd}

\begin{widetext}
\section*{Supplementary Material}
\setcounter{section}{0}
\setcounter{equation}{0}
\setcounter{figure}{0}
\renewcommand*{\thefigure}{S\arabic{figure}}
\renewcommand*{\theequation}{S\arabic{equation}}
\renewcommand\thesection{Supplementary Note \Roman{section}}
\section{Leaked Information Calculation}\label{sec-S1}
Under the asymptotic limit and heterodyne-detection conditions, the leaked information ${\chi _{{\rm{bE}}}}$ in the GMCS-CVQKD system is calculated by the following formula:
\begin{align}
	{\chi _{{\rm{bE}}}} =& \sum\limits_{i = 1}^2 {G\left( {\frac{{{\lambda _i} - 1}}{2}} \right)}  - \sum\limits_{i = 3}^5 {G\left( {\frac{{{\lambda _i} - 1}}{2}} \right)} ,\label{Eq.S1}\\
	{\lambda _{1,2}} =& \sqrt {\frac{1}{2}\left( {A \pm \sqrt {{A^2} - 4B} } \right)}, {\lambda _{3,4}} = \sqrt {\frac{1}{2}\left( { \pm \sqrt {{C^2} - 4D} } \right)} ,{\lambda _5} = 1, \label{Eq.S2}	\\
	A =& {V^2}\left( {1 - 2T} \right) + 2T + {T^2}{\left( {V + {\chi _{{\rm{line}}}}} \right)^2}, 
		B = {T^2}{\left( {V{\chi _{{\rm{line}}}} + 1} \right)^2},\label{Eq.S3}\\
	C=&\frac{A\chi_{\mathrm{het}}^2+B+1
		+2\chi_{\mathrm{het}}(V\sqrt{B}+T(V+\chi_{\mathrm{line}}))+2T(V^2-1)}{(T(V+\chi_{\mathrm{tot}}))^2},D = {\left( {\frac{{V + \sqrt B {\chi _{{\rm{het}}}}}}{{T\left( {V + {\chi _{{\rm{tot}}}}} \right)}}} \right)^2},\label{Eq.S4}
\end{align}
where $G\left( x \right) = \left( {x + 1} \right){\log _2}\left( {x + 1} \right) - x{\log _2}\left( x \right),V = {V_A} + 1,{\chi _{{\rm{line}}}} = 1/T - 1 + \varepsilon $ denotes the channel noise at the channel input. ${\chi _{{\rm{het}}}} = \left[ {1 + \left( {1 - \eta } \right) + 2{v_{{\rm{el}}}}} \right]/\eta $ denotes the detection noise referred to Bob's input. ${\chi _{{\rm{tot}}}} = {\chi _{{\rm{line}}}} + {\chi _{{\rm{het}}}}/T$ represents the total noise referred to the channel input.

\section{Post-Processing Procedure for Experimental Data}\label{sec-S2}
The post-processing procedure of the experimental data is described below. First, the peak searching and downsampling are required for the oversampled detection results. Since the oscilloscope used for data acquisition and the signal generator are triggered by the same clock, subsequent downsampling can be performed by simply identifying the peak point of each pulse. Considering that the pilot signal has relatively high energy, its peak position can be located more precisely, while the final time delay of the quantum signal relative to the pilot signal can be calibrated in advance. Therefore, once the peak points of the pilot signal are determined, the peak points of the quantum signal can be determined accordingly. Given a sampling rate of 2.5 Gsample/s and a symbol rate of 50 MHz, corresponding to an oversampling factor of 50, the power
${P_i} = x_{Bi}^2 + p_{Bi}^2$ of all pilot signals can be calculated. Furthermore, data are extracted every 50 points using different sampling positions as starting points, and the corresponding energy values are compared. The sampling position with the maximum energy is then selected as the downsampling point.

After downsampling, the center wavelengths of the LO and the signal laser are first roughly calibrated using an optical spectrum analyzer, while the remaining center wavelength deviation is compensated through frequency-offset estimation. Specifically, the pilot signal ($X_P+iP_P$) is transformed into the frequency domain using the Fourier transform, and the frequency $f_{\mathrm{off}}$ corresponding to the maximum spectral component is identified as the frequency offset. Subsequently, the frequency offset recovery can be performed using the following formula,
\begin{align}
	X_P \leftarrow& \Re{((X_P+iP_P)e^{-i2\pi f_{\mathrm{off}}t})}, 	P_P \leftarrow \Im{((X_P+iP_P)e^{-i2\pi f_{\mathrm{off}}t})},\label{Eq.S5}\\
	X_Q \leftarrow& \Re{((X_Q+iP_Q)e^{-i2\pi f_{\mathrm{off}}t})},
	 P_Q \leftarrow \Im{((X_Q+iP_Q)e^{-i2\pi f_{\mathrm{off}}t})},\label{Eq.S6}
\end{align}

The phase drift of the quantum signal needs to be further compensated after frequency offset recovery. Since the pilot signal and the quantum signal pulses are generated synchronously, the phase of the pilot signal can be used as a reference to correct the fast phase drift of the quantum signal,
\begin{align}
	{\theta ^{{\rm{rot}}}} = \frac{{{X_P} + i{P_P}}}{{\left| {{X_P} + i{P_P}} \right|}},{X_Q} + i{P_Q} \leftarrow \frac{{{X_Q} + i{P_Q}}}{{{\theta ^{{\rm{rot}}}}}},\label{Eq.S7}
\end{align}
where $\theta^{\mathrm{rot}}$ represents the phase of the pilot signal, $X_Q$ and $P_Q$ represent the $X$ and $P$ quadratures of the quantum signal, respectively.

Finally, to achieve frame synchronization, the first 50\% of the data in each frame is disclosed as a frame header, and the correlation between the frame header and the received data is calculated. The number of cyclic shifts is determined based on the peak of the correlation calculation, thereby completing frame synchronization. Subsequently, the slow drift in the received data is compensated. At this point, the entire post-processing procedure is completed.

\begin{figure}[t]
	\centering
	\includegraphics[width=\linewidth]{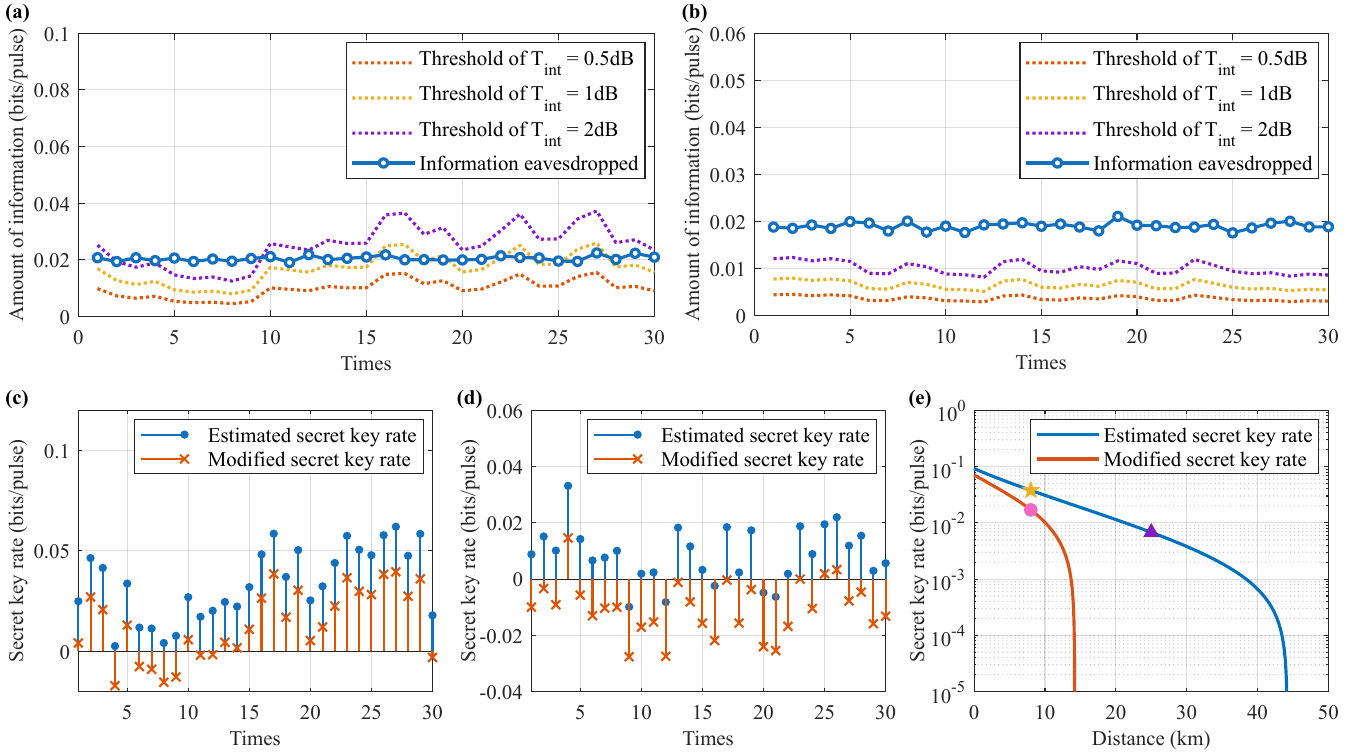}
	\caption{Attacking results with cutoff frequency 150 MHz. Comparison of the information obtained by Eve via spectral attack with the minimum information required to render the system insecure at (a) 8 km and (b) 25km; Secret key rates obtained from parameter estimation and modified secret key rates after considering spectral attack at (c) 8 km and (d) 25 km; (e) Secret key rates obtained from parameter estimation and modified secret key rates as a function of distance. The secret key rate curves are simulated based on average values for the estimated experimental parameters of each frame. The yellow pentagon, red circle and purple triangle represent experimental points at 8 km, 8km and 25 km. The modified secret key rate at 25 km is negative and is not plotted.}\label{fig-8}
\end{figure}

\section{Experimental Results at a Cutoff Frequency of 150 MHz}\label{sec-S3}
In this section, we give out the experimental results at a cutoff frequency of 150 MHz. Fig.~\ref{fig-8}(a) and (b) show the comparison between the amount of information eavesdropped by Eve and the minimum information required to render the system insecure at transmission distances of 8 km and 25 km, respectively. The blue curve represents the amount of information actually obtained by Eve over 30 experiments,  while the three dashed lines correspond to the minimum information required to render the system insecure when the wavelength splitter insertion loss is 0.5 dB, 1 dB, and 2 dB, respectively. Under conditions of low insertion loss, Eve is able to acquire sufficient information in certain experiments to compromise the security of the system. As the insertion loss increases, the conditions for a successful attack become significantly more stringent. In particular, at an insertion loss of 2 dB, the information obtained by Eve does not exceed the threshold in the majority of experiments. At a transmission distance of 25 km, the amount of information actually obtained by Eve exceeds the corresponding threshold at all experimental points. This indicates that, as the transmission distance increases, the tolerance of the system to additional information leakage decreases. Consequently, a smaller amount of eavesdropped information is sufficient to render the system insecure. This is consistent with the results presented in the main text.

Fig.~\ref{fig-8}(c) and (d) show the system secret key rates at transmission distances of 8 km and 25 km, respectively. The blue data represent the original secret key rate obtained by the legitimate parties through conventional parameter estimation, while the orange data represent the actual secret key rate modified by considering the spectral attack. It can be observed that the modified secret key rate is lower than the original secret key rate at all experimental points, indicating that neglecting out-of-band information leakage would lead the legitimate parties to overestimate the security of the system. At a transmission distance of 25 km, the modified secret key rate decreases more noticeably, reflecting the greater impact of the spectral attack on system security over longer distances. Fig.~\ref{fig-8}(e) illustrates the trend of the secret key rate as channel loss. While both the original estimated secret key rate and the modified secret key rate decrease as the channel loss increases, the modified secret key rate decreases more rapidly and drops to zero at relatively low channel loss. This suggests that the spectral attack significantly narrows the actual security region of the system, further highlighting the importance of considering out-of-band information leakage in security estiamtion.

By comparing the results under cutoff frequencies of 100 MHz and 150 MHz, it can be observed that the actual amount of information obtained by Eve is generally higher at 100 MHz than at 150 MHz. Specifically, at a transmission distance of 8 km, most experimental points exceed the minimum information threshold required to render the system insecure. In contrast, under the 150 MHz condition, the information available to Eve is reduced, making the conditions for successful attack implementation more stringent. Furthermore, under the 100 MHz cutoff frequency, the modified secret key rate decreases more significantly relative to the original estimate. Under the 150 MHz condition, although the modified key rate is still lower than the original estimate, the reduction is relatively smaller, suggesting that the impact of the attack on the system’s security boundary is less pronounced. As the cutoff frequency increases, Bob can cover more effective frequency components that would otherwise leak into the out-of-band region, thereby reducing the out-of-band information available to Eve. These results demonstrate that appropriately increasing the effective bandwidth at the receiver and reducing the bandwidth mismatch can help suppress out-of-band information leakage and decrease the feasibility of the spectral attack. However, under long-distance transmission conditions, the system remains sensitive to additional information leakage even when the cutoff frequency is increased. Therefore, the impact of bandwidth configuration on system security must still be carefully considered in the design of practical CVQKD systems.

\end{widetext}

\end{document}